\documentclass
[12pt]{article}
\usepackage{amssymb}
\usepackage{epsfig}

\catcode`\@=11
\def\numberbysection{\@addtoreset{equation}{section}
        \def\theequation{\thesection.\arabic{equation}}}

\def\beq{\begin{equation}}
\def\eeq{\end{equation}}
\numberbysection
\begin{document}
\begin{titlepage}
\begin{center}
\hfill  \\
\vskip 1.in {\Large \bf Dirac oscillator and minimal length} \vskip 0.5in P. Valtancoli
\\[.2in]
{\em Dipartimento di Fisica, Polo Scientifico Universit\'a di Firenze \\
and INFN, Sezione di Firenze (Italy)\\
Via G. Sansone 1, 50019 Sesto Fiorentino, Italy}
\end{center}
\vskip .5in
\begin{abstract}
We show how to solve the Dirac oscillator with a minimal length by using previous results on the harmonic oscillator in a Snyder algebra.
\end{abstract}
\medskip
\end{titlepage}
\pagenumbering{arabic}
\section{Introduction}

It is expected that a consistent formulation of quantum gravity requires the introduction of a minimal length scale of the order of the Planck scale.

It is well known that a standard formulation of quantum mechanics with a minimal length is obtained by deforming the commutation relations between position and momentum operator (GUP) \cite{1}. This formulation is solvable in practice only in a limited number of physical systems, such as for example the harmonic oscillator in $d$ dimensions and the Dirac oscillator.

The purpose of this work is to investigate the problem of quantization of the non-commutative Dirac oscillator. Although this problem has already been discussed in some works
 \cite{2} - \cite{3}, we want to follow an independent method, based on the solution of the harmonic oscillator in $d$ dimensions that we developed in \cite{4}.

 The article is organized as follows. Firstly we remember the eigenvalue problem for a harmonic oscillator in $d$ dimensions with a minimal length, giving a more precise characterization
of the eigenvectors of the problem. Then in Section 4 we are able to accurately link the problem of quantization of the Dirac oscillator
to that of the harmonic oscillator in 3 dimensions for both two-component spinors of the Dirac equation.

We are also able to extend the non-commutative deformation of the Heisenberg algebra from the Snyder algebra ($ \beta $ quantization) to the case of a deformation in two parameters (($ \alpha, \beta $) quantization \cite{5} ) and discuss how the spectrum of eigenvalues is changed also in this case.

\section{Harmonic oscillator revisited}

Let us recall the method with which we have solved the problem of quantization of the harmonic oscillator in $d$ dimensions with the Snyder algebra. The Hamiltonian is defined by

\beq H \ = \ \sum^{d}_{i=1} \ \left( \frac{p^2_i}{2m} \ + \ \frac{1}{2} m \omega^2 x^2_i \right) \label{21} \eeq

with the following choice of the commutation rules

\begin{eqnarray}
& \ & [ x_i, p_j ] \ = \ i \hbar \  ( \delta_{ij} \ + \ \beta p_i p_j ) \nonumber \\
& \ & [ p_i, p_j ] \ = \ 0 \nonumber \\
& \ & [ x_i, x_j ] \ = \ i \hbar \beta \ ( p_j x_i \ - \ p_i x_j ) \label{22}
\end{eqnarray}

also called Snyder algebra. This set of commutation rules is resolved by the following representation in the compact variable $\rho$

\beq x_i \ = \ i \hbar \ \sqrt{ 1 -\beta\rho^2 } \ \frac{\partial}{\partial\rho_i} \ \ \ \ \ \ \ \ \ \ \ p_i \ = \ \frac{\rho_i}{\sqrt{ 1-\beta\rho^2 }} \ \ \ \ \ \ \ \
0 < \rho^2 < \frac{1}{\beta}
\label{23} \eeq

By inserting this representation in the Hamiltonian, the eigenvalue problem is defined by

\beq \sum^d_{i=1} \ \left( \frac{\rho^2_i}{1-\beta\rho^2} \ - \ \frac{1}{\beta^2 \mu ( \mu-1 )} \ \sqrt{ 1-\beta\rho^2 } \ \frac{ \partial }{ \partial \rho_i } \ \sqrt{ 1-\beta\rho^2 } \ \frac{ \partial }{ \partial \rho_i }
\right) \ \psi( \rho ) \ = \ 2 m E \ \psi ( \rho ) \label{24} \eeq

where the collective $\mu$ parameter is tied to the basic parameters of the theory

\beq \frac{1}{\sqrt{\mu ( \mu - 1)}} \ = \ m \omega \hbar \beta \label{25} \eeq

The ground state is easily identifiable as

\beq \psi_0 ( \rho ) \ = \ c_0 \ {( 1-\beta\rho^2 )}^{\frac{\mu}{2}} \label{26} \eeq

with associated eigenvalue

\beq E_0 \ = \ d \ \frac{\hbar \omega}{2} \ \sqrt{\frac{\mu}{\mu-1}} \label{27} \eeq

To discuss the excited states, we introduce the following parametrization of the wave function

\beq \psi ( \rho ) \ = \ \chi ( \rho ) \ {( 1-\beta\rho^2 )}^{\frac{\mu}{2}} \label{28} \eeq

and defining new variables $z_i \ = \ \sqrt{ \beta } \rho_i$, we obtain the differential equation for $\chi(\rho)$

 \beq \left[ \ [ 1 - ( \sum^d_{i=1} z^2_i ) ] \ \sum^d_{i=1} \ \frac{\partial^2}{\partial z_i \partial z_i} \ - \ ( 1+2\mu ) \ \sum^d_{i=1} z_i \frac{\partial}{\partial z_i} \ + \ \epsilon_{\{n_i\}} \ \right] \ \chi ( \rho ) \ = \ 0 \label{29} \eeq

where

 \beq E_{\{n_i\}} \ = \ \frac{\hbar \omega}{2} \ \frac{ ( \ \epsilon_{\{n_i\}} \ + \ d \  \mu \ ) }{\sqrt{\mu ( \mu-1 )}} \label{210} \eeq

the energy parameter depends on some quantum numbers $\{n_i\}$.

The eigenvalue problem is solved by looking for polynomial solutions in $\chi(\rho)$

\beq \chi(\rho) \ = \ P_{\{n_i\}} ( z_i) \label{211} \eeq

From the discussion in \cite{4} we derive the general formula for $ \epsilon_{\{n_i\}} $

\beq \epsilon^{\mu,d}_{N,l} \ = \ N ( 1 + 2 \mu ) + ( N - l ) ( N + l + d -2 ) \label{212} \eeq

where $ N = n + l $ and $n$ is a positive even integer. Then the spectrum of the eigenvalues depends on two quantum numbers $N$, the total quantum number, and $l$ related to the angular momentum.

In this article we also give a more precise characterization of the eigenvectors. The wave function can be decomposed as

\beq \psi_{n,l}^{\{m_i\}} ( \rho_i ) \ = \ c \ P^{\mu,d}_{n,l} (r) \ Y_l^{\{m_i\}} ( \theta_i) \ ( 1 - r^2 )^\frac{\mu}{2} \label{213} \eeq

where $ r = \sqrt{| z_i z_i |} $ and the angular part is solved by the generalized spherical harmonics in $d$ dimensions.

The polynomial $ P^{\mu,d}_{n,l} (r) $ can be defined as

\beq P^{\mu,d}_{n,l} (r) \ = \ r^l \ \sum^\frac{n}{2}_{k = 0} \ a_k ( \mu ) \ r^{2k} \label{214} \eeq

where the constants $ a_k $ are determined by the eigenvalue equation (\ref{29}) with the following recurrence equation:

\begin{eqnarray} a_0 & = & 1 \nonumber \\ a_{k+1} & = & - \ a_{k} \ \frac{( n - 2 k )( n + 2\mu + 2 k + 2 l + d - 1 )}{( 2 k + 2 )( 2 k + 2 l + d )} \label{215} \end{eqnarray}

The polynomial $ P^{\mu,d}_{n,l} (r) $ satisfies the equation

\beq \nabla^2 \ P^{\mu,d}_{n,l} (r) \ = \ - \ ( \epsilon^{\mu,d}_{n,l} \ - \ l ( 2 \mu + 1 ) ) \ P^{\mu+2,d}_{n-2,l} (r) \label{216} \eeq

where the differential operator $ \nabla^2 $ in spherical coordinates is defined by

\beq  \nabla^2 \ = \ \frac{d^2}{dr^2} \ + \ \frac{d-1}{r} \ \frac{d}{dr} \ - \ \frac{l(l+d-2)}{r^2} \label{217} \eeq

and

\beq \epsilon^{\mu,d}_{N,l} \ = \ ( n + l )  ( 1 + 2 \mu ) + n ( n + 2l + d -2 ) \label{218} \eeq

Despite the recurrence formula (\ref{215}) is complicated, the normalization of the polynomials is relatively simple

\beq \int^1_0 \ dr \ r^{d-1} \ [ P^{\mu,d}_{n,l} (r) ]^2 \ ( 1-r^2 )^{\mu - \frac{1}{2}}  \ = \ N^{\mu,d}_{n,l} \label{219} \eeq

where

\beq N^{\mu,d}_{n,l} \ = \ \frac{\Gamma(\frac{d}{2})}{2^{l+1}} \ \frac{((2l+d-2)!!)^2 \ n!!}{(d-2)!! \ (n+2l+d-2)!!} \frac{\Gamma(\frac{1+n}{2} + \mu)}{(\mu + n + l + \frac{d-1}{2}) \Gamma(\mu + \frac{n}{2} + l + \frac{d-1}{2}) } \label{220} \eeq

a formula valid for each $n$, $l$ and  $d$.

For the special case $d=2$ it simplifies to

\beq N^{\mu,2}_{n,l} \ = \ \frac{(2^l l!)^2 \ \prod^l_{i=1} \left( \frac{1}{n+2i}\right)}{(2\mu +2(n+l)+1) \ \prod^{l-1}_{i=0} ( 2 \mu + n + 2i + 1 )}  \label{221} \eeq

and for $ d = 3 $

\beq N^{\mu,3}_{n,l} \ = \ \frac{\sqrt{\pi}}{2^{l+2}} \ \frac{((2l+1)!!)^2 \ n!!}{  (n+2l+1)!!} \frac{\Gamma(\frac{1+n}{2} + \mu)}{(\mu + n + l + 1) \Gamma(\mu + \frac{n}{2} + l + 1 ) } \label{222} \eeq

These formulas have been obtained with the help of MATHEMATICA and have been verified in several cases.

\section{($ \alpha, \beta $) quantization}

We can easily extend the results of section 2  in the case of the two parameter non-commutative quantization ($\alpha$, $\beta$) defined by

\begin{eqnarray} & \ & [ x_i, p_j ]  =  i \hbar ( \delta_{ij} + \alpha x_i x_j + \beta p_j p_i + \sqrt{\alpha \beta} ( p_i x_j + x_j p_i ) ) \nonumber \\
& \ & [ x_i, x_j ] =  i \hbar \beta \epsilon_{ijk} L_k \nonumber \\
& \ & [ p_i, p_j ] =  i \hbar \alpha \epsilon_{ijk} L_k \nonumber \\
&\ & \epsilon_{ijk} L_k = \frac{1}{2} ( x_i p_j + p_j x_i - x_j p_i - p_i x_j ) \label{31}
\end{eqnarray}

This generalized commutation rule can always be resolved with a $\rho$ parametrization

\begin{eqnarray}
x_i & = & i \hbar \sqrt{1-\beta\rho^2} \frac{\partial}{\partial \rho_i} \ + \ \lambda \sqrt{\frac{\beta}{\alpha}} \ \frac{\rho_i}{\sqrt{1-\beta\rho^2}} \nonumber \\
p_i & = & - i \hbar \sqrt{\frac{\alpha}{\beta}} \sqrt{1-\beta\rho^2} \frac{\partial}{\partial \rho_i} \ + \ ( 1 - \lambda ) \frac{\rho_i}{\sqrt{1-\beta\rho^2}} \label{32}
\end{eqnarray}

with $\lambda$ an arbitrary parameter.

In this article we want to point out that it is possible to map the problem of quantization of the harmonic oscillator with parameters ($\alpha, \beta$) in terms of the quantization of the harmonic oscillator with only one parameter $\beta$ ( Snyder algebra ) with an appropriate choice of the parameter $\lambda$. First we decompose the general representation ($ x_i, p_i $) in terms of a representation of the Snyder algebra ($ x_i^0, p_i^0 $)

\begin{eqnarray}
x_i & = & x_i^0 \ + \ \lambda \sqrt{\frac{\beta}{\alpha}} \ p_i^0 \nonumber \\
p_i & = & - \sqrt{\frac{\alpha}{\beta}} \ x_i^0 \ + \ ( 1 - \lambda ) \ p_i^0  \label{33}
\end{eqnarray}

Therefore the term of harmonic oscillator in the variables ($x_i, p_i$)

\begin{eqnarray}
p_i^2 + m^2 \omega^2 x_i^2  & = & \left( \frac{\alpha}{\beta} + m^2 \omega^2  \right) ( x_i^0 )^2 \ + \ \left( ( 1 - \lambda )^2 + m^2 \omega^2 \lambda^2 \frac{\beta}{\alpha} \right)
( p_i^0 )^2 + \nonumber \\ & + & \left( m^2 \omega^2 \lambda \sqrt{\frac{\beta}{\alpha}} - ( 1-\lambda ) \sqrt{\frac{\alpha}{\beta}} \right) ( x_i^0 p_i^0 + p_i^0 x_i^0 ) \label{34}
\end{eqnarray}

can be brought to an harmonic oscillator in the variables ($x_i^0, p_i^0$ ) by choosing

\beq \lambda = \frac{\alpha}{\alpha + m^2 \omega^2 \beta} \label{35} \eeq

from which

\beq  ( p_i^2 + m^2 \omega^2 x_i^2 ) \psi = \frac{1}{h} \left[  ( p_i^0 )^2 + m^2 \omega^2 h^2  ( x_i^0 )^2  \right] \psi = 2 m E \psi \label{36} \eeq

where the factor $h$ is defined by

\beq h = 1 + \frac{\alpha}{\beta m^2 \omega^2} \label{37} \eeq

The factor $h$ can be absorbed by a redefinition of the mass

\beq m \rightarrow m' = m h = m \left( 1 + \frac{\alpha}{\beta m^2 \omega^2} \right) \label{38} \eeq

Then the spectrum of the eigenvalues is identical to the case of the quantization of the Snyder algebra provided the dependence of the parameter $\mu$ from the basic parameters
of the theory is modified as follows

\beq \frac{1}{\sqrt{\mu ( \mu -1 )}} \ = \ m' \omega \hbar \beta \ = \ \hbar \left( m \omega \beta + \frac{\alpha}{m \omega} \right) \label{39}\eeq

\section{Dirac oscillator}

Our main interest is the solution of the Dirac oscillator in a Snyder algebra in which case we introduce the matrix equation:

\beq [ \ \tilde{\alpha}_i \ ( p_i  - i \tilde{\beta} \  m \omega x_i ) \ + \ \tilde{\beta} \ m  \ ] \psi  \ = \ E \psi \label{41}\eeq

where

\beq \psi = \left( \begin{array}{c} \psi_a \\ \psi_b \end{array} \right)  \label{42}\eeq

is a four component spinor that can be decomposed in the two-component spinors $\psi_a$ and $\psi_b$ and the matrices $\tilde{\alpha}_i$ and $\tilde{\beta}$ are defined as usual as

\beq \tilde{\alpha}_i = \left( \begin{array}{cc} 0 & \sigma_i \\ \sigma_i & 0  \end{array}
\right) \ \ \ \ \ \ \ \ \ \ \ \ \ \tilde{\beta} = \left( \begin{array}{cc} 1 & 0 \\ 0 & - 1  \end{array} \right)  \label{43} \eeq

The eigenvalue equation can be decomposed into two coupled equations

\begin{eqnarray} ( E-m ) \ \psi_a \ = \ ( \ \sigma_i p_i \ + \ i m \omega \sigma_i x_i \ ) \ \psi_b \nonumber \\
( E+m ) \ \psi_b \ = \ ( \ \sigma_i p_i \ - \ i m \omega \sigma_i x_i \ ) \psi_a \label{44} \end{eqnarray}

Resolving with respect to the $ \psi_a $ spinor we obtain the quadratic equation

\begin{eqnarray} ( E^2 -m^2 ) \ \psi_a( p_i ) & = & [ \ p_i^2 + m^2 \omega^2 x_i^2 + i m \omega [ \sigma_i x_i , \sigma_j p_j ] + \nonumber \\
& + & i m^2 \omega^2 \epsilon_{ijk} \sigma_i x_j x_k ] \ \psi_a ( p_i )
\label{45} \end{eqnarray}

where the commutators can be easily calculated

\begin{eqnarray}  & \ & \epsilon_{ijk} \  x_j x_k  \ = \ i \hbar \beta L_i \nonumber \\
& \ & [ \sigma_i x_i , \sigma_j p_j ] \ = \ i \hbar \left( 2 \frac{\sigma_i L_i}{\hbar} + \beta p_i^2 + 3 \right)
\label{46} \end{eqnarray}

from which the equation for $\psi_a$ can be rewritten as

\begin{eqnarray} ( E^2 -m^2 ) \ \psi_a( p_i ) & = & [ \ p_i^2 ( 1 - m \omega \hbar \beta ) \ + \ m^2 \omega^2 x_i^2 - \nonumber \\
& - & m \omega ( 2 + m \omega \hbar \beta  ) \sigma_i L_i \ - \ 3 m \omega \hbar \ ] \psi_a ( p_i )
\label{47} \end{eqnarray}

The part similar to the harmonic oscillator

\beq [ \ p_i^2 ( 1 - m \omega \hbar \beta ) \ + \ m^2 \omega^2 x_i^2 \ ] \psi_a \ = \ 2 m W^{(1)} \ \psi_a
\label{48} \eeq

leads to an equation similar to the case discussed in the second section provided the relationship between $\mu$ and the combination $m  \omega \hbar \beta$ is changed:

\beq \frac{1}{\mu} \ = \ m \omega \hbar \beta \ \ \ \ \ \ \ \ \ \ \ \ \psi_a \ \ {\rm spinor} \label{49} \eeq

The ground state of $\psi_a$ is therefore

\beq \psi_a^0 ( \rho_i ) \ = \ c_0 \ ( 1 - \beta \rho^2 )^{\frac{\mu}{2}} \label{410} \eeq

with energy

\beq W^{(1)}_0 \ = \ \frac{3}{2} \ \hbar \omega \label{411} \eeq

In general for the excited states it is obtained

\begin{eqnarray}  & \ &  W^{(1)} =  \frac{\hbar \omega}{2} \ \left[ \  \frac{ \epsilon_{\{n_i\}} + 3 \ \mu }{ \mu } \
\right] \nonumber \\
& \ & \epsilon_{\{n_i\}} =  N ( 1 + 2 \mu ) \ + \ ( N - l ) ( N + l + 1 )
\label{412} \end{eqnarray}

As for the spin-orbit interaction (which is a constant in the variables $(x_i, p_i)$) and the constant interaction we obtain

\begin{eqnarray}  & \ & [ \ - m \omega \ ( 2 + m \omega \hbar \beta ) \ \sigma_i L_i \ - 3 m \omega \hbar \ ] \ \psi_a \ = \ 2 m W^{(2)} \ \psi_a \nonumber \\
& \ & \sigma_i L_i \ = \ \hbar \left( \ J^2 - L^2 - \frac{3}{4} \ \right) \ = \ \hbar \left( \ j(j+1) - l ( l+1 ) - \frac{3}{4} \ \right)
\label{413} \end{eqnarray}

where $ j = l \pm \frac{1}{2} $. Therefore we obtain for $ W^{(2)} $:

\beq W^{(2)} \ = \ - \hbar \omega \left( \ 1 + \frac{1}{2\mu} \right) \ \left( \ j(j+1) - l ( l+1 ) - \frac{3}{4} \right) \ - \frac{3}{2} \ \hbar \omega
\label{414} \eeq

The final eigenvalue is the sum of $W^{(1)}$ and $W^{(2)}$

\begin{eqnarray} W_a & = & \frac{ E^2-m^2 }{2m} \ = \ \frac{\hbar \omega}{2} \ \left[ \ \frac{ N ( 1 + 2 \mu ) + ( N - l ) ( N + l + 1 )  + 3 \mu }{ \mu } \
\right] - \nonumber \\
& - & \hbar \omega \left( \ 1 + \frac{1}{2\mu} \right) \ \left( \ j(j+1) - l ( l+1 ) - \frac{3}{4} \ \right) - \frac{3}{2} \ \hbar \omega \label{415} \end{eqnarray}

This formula concludes the discussion for the component $\psi_a$.

Repeating the same argument for the component $\psi_b$ we obtain

\begin{eqnarray} ( E^2 -m^2 ) \ \psi_b( p_i ) & = & [ \ p_i^2 ( 1 + m \omega \hbar \beta ) \ + \ m^2 \omega^2 x_i^2 + \nonumber \\
& + & m \omega ( 2 - m \omega \hbar \beta  ) \sigma_i L_i \ + \ 3 m \omega \hbar \ ] \psi_b ( p_i )
\label{416} \end{eqnarray}

The part analogous to the harmonic oscillator

\beq [ \ p_i^2 ( 1 + m \omega \hbar \beta ) \ + \ m^2 \omega^2 x_i^2 \ ] \psi_b \ = \ 2 m W^{(1)} \ \psi_b
\label{417} \eeq

can be linked to the standard harmonic oscillator case with the substitution of the equation (\ref{25}) with

\beq \frac{1}{\mu - 1} \ = \ m \omega \hbar \beta \ \ \ \ \ \ \ \ \ \ \ \ \psi_b \ \ {\rm spinor} \label{418} \eeq

The ground state of $\psi_b$ is

\beq \psi_b^0 ( \rho_i ) \ = \ c_0 \ ( 1 - \beta \rho^2 )^{\frac{\mu}{2}} \label{419} \eeq

with energy

\beq W^{(1)}_0 \ = \ \frac{3}{2} \ \hbar \omega \frac{\mu}{\mu - 1} \label{420} \eeq

For the spectrum in general we obtain

\begin{eqnarray}  & \ &  W^{(1)} =  \frac{\hbar \omega}{2} \ \left[ \  \frac{ \epsilon_{\{n_i\}} + 3 \mu }{ \mu - 1 } \
\right] \nonumber \\
& \ & \epsilon_{\{n_i\}} =  N ( 1 + 2 \mu ) \ + \ ( N - l ) ( N + l + 1 )
\label{421} \end{eqnarray}

Similarly for the spin-orbit interaction and the constant interaction we get the contribution

\beq W^{(2)} \ = \ \hbar \omega \left( \ 1 - \frac{1}{2 ( \mu-1 ) } \right) \ \left( \ j(j+1) - l ( l+1 ) - \frac{3}{4} \ \right) \ + \ \frac{3}{2} \ \hbar \omega
\label{422} \eeq

from which the final eigenvalue is

\begin{eqnarray} W_b & = & \frac{ E^2-m^2 }{2m} \ = \ \frac{\hbar \omega}{2} \ \left[ \ \frac{ N ( 1 + 2 \mu ) + ( N - l ) ( N + l + 1 )  + 3 \mu }{ \mu - 1 } \
\right] + \nonumber \\
& + & \hbar \omega \left( \ 1 - \frac{1}{2 ( \mu-1 ) } \right) \ \left( \ j(j+1) - l ( l+1 ) - \frac{3}{4} \ \right) + \frac{3}{2} \ \hbar \omega \label{423} \end{eqnarray}

\section{Dirac oscillator with ($ \alpha, \beta $) quantization }

The quantization of the Dirac oscillator with noncommutative parameters ($\alpha, \beta$) is obtained ( as described in section 3 ) by replacing

\begin{eqnarray}
x_i & = & x_i^0 \ + \ \lambda \sqrt{\frac{\beta}{\alpha}} \ p_i^0 \nonumber \\
p_i & = & - \sqrt{\frac{\alpha}{\beta}} \ x_i^0 \ + \ ( 1 - \lambda ) \ p_i^0  \label{51}
\end{eqnarray}

where it is convenient to choose $ \lambda = \frac{\alpha}{\alpha + m^2 \omega^2 \beta} $ and the variables ( $ x_i^0 , p_i^0 $ ) satisfy the Snyder algebra.

For the component $ \psi_a $ we get

\begin{eqnarray} & \ & ( \sigma_i p_i + i m \omega \sigma_i x_i ) ( \sigma_j p_j - i m \omega \sigma_j x_j ) \ = \ \frac{m^2 \omega^2 \beta}{\alpha + m^2 \omega^2 \beta} \ (p_i^0)^2 +
\nonumber \\ & \ & + \left( \frac{\alpha}{\beta} + m^2 \omega^2 \right) ( \ ( x_i^0 )^2 + i \epsilon_{ijk} \sigma_i x^0_j x^0_k \ ) + i m \omega [ \sigma_i x_i^0, \sigma_j x_j^0 ] \label{52}
\end{eqnarray}

The part analogous to the harmonic oscillator can be written as

\beq \frac{1}{h} \  [ \ (p_i^0)^2 ( 1 - h m \omega \hbar \beta ) + m^2 \omega^2 h^2 ( x_i^0 )^2  \ ] \psi_a \ = \ 2 m W^{(1)} \ \psi_a \label{53} \eeq

and can be linked to the standard case (\ref{24}) of the harmonic oscillator with the choice of the parameter $\mu$

\beq \frac{1}{\mu} \ = \ h  m \omega \hbar \beta  \ = \ \hbar \left( \beta m \omega + \frac{\alpha}{m \omega}
\right) \ \ \ \ \ \ \ \ \ \ \ \psi_a \ \ {\rm spinor} \label{54} \eeq

All the above discussion is identical with the substitution of (\ref{54}) in the final formula (\ref{415}).

Similarly for the component $\psi_b$ we need to modify only the choice of the collective parameter $\mu$:

\beq \frac{1}{ \mu - 1 } \ = \ h  m \omega \hbar \beta  \ = \ \hbar \left( \beta m \omega + \frac{\alpha}{m \omega}
\right) \ \ \ \ \ \ \ \ \ \ \ \psi_b \ \ {\rm spinor} \label{55} \eeq

All formulas (\ref{421}), (\ref{422}) and (\ref{423}) are still valid with the replacement of (\ref{55}) instead of (\ref{418}).

\section{Conclusion}

In this article we have shown how to solve the problem of quantization of a $3d$ Dirac oscillator with minimal length. Although the equations of the classical Dirac oscillator are relativistic, the similar non-commutative ones  aren't, because the non-commutative deformation we discuss here treats the variables ($ t, $ E) as classical and the variables ($ x_i, p_i $)
as non-commutative. The solution we propose is based on the polynomial solutions of a differential equation (\ref{29}) we have discussed for the first time in \cite{4}. These solutions are a $d$-dimensional generalization of the Gegenbauer polynomials. In this article we have discussed the algebraic properties of these new polynomials, obtained with the help of MATHEMATICA.

We have also been able to connect the quantization of the Dirac oscillator with the differential equation (\ref{29}). The main difference with the harmonic oscillator is the choice of the collective parameter $\mu$, which controls the polynomials, compared to the basic parameters of the theory. Our solution is elegant and more simple compared to the other articles written on the same subject \cite{2}-\cite{3}.


\begin{thebibliography}{999}
\bibitem{1} A. Kempf, G. Mangano, R. B. Mann,  Phys. Rev. {\bf D 52} (1995) 1108, hep-th/9412167.
\bibitem{2} M. M. Stetsko, J.Math.Phys. {\bf 56} (2015) 012101, arXiv:1310.0706.
\bibitem{3} M. Betrouche, M. Maamache, J. R. Choi, Advances in High Energy Physics {\bf 2013}, Article ID 383957.
\bibitem{4} P. Valtancoli, Prog. Theor. Exp. Phys. {\bf 2014} 023A02, arXiv:1306.0116.
\bibitem{5} S. Mignemi, Class.Quant.Grav. {\bf 29} (2012) 215019, arXiv:1110.0201.
\end{thebibliography}
\end{document}